\documentclass{iau}
\usepackage{graphicx}

\title[A radiative-convective equilibrium model for young giant exoplanets] 
{A radiative-convective equilibrium model for young giant exoplanets: \\ Application to $\beta$ Pictoris b}

\author[Jean-Loup Baudino \etal]   
{Jean-Loup Baudino$^1$
 , Bruno B\'ezard$^1$
 , Anthony Boccaletti$^1$
 , Mickael Bonnefoy$^2$
 \and Anne-Marie Lagrange$^3$}

\affiliation{$^1$LESIA, Observatoire de Paris, CNRS, UPMC, Universit\'e Paris-Diderot, 5 place Jules Janssen,\\ F-92195 Meudon, France \\ email: {\tt jean-loup.baudino@obspm.fr} \\[\affilskip]
$^2$Max Planck Institut of Astronomy, K\"onigstuhl17,\\ D-69117 Heidelberg, Germany
\\[\affilskip]
$^3$UJF-Grenoble 1 / CNRS-INSU, Institut de Plan\'etologie et d'astrophysique de Grenoble (IPAG) UMR 5274,\\ Grenoble, F-38041, France}

\pubyear{2013}
\volume{299}  
\pagerange{}
\jname{Exploring the Formation and Evolution of Planetary
Systems}
\editors{Brenda Matthews \& James Graham}
\begin{document}

\maketitle

\begin{abstract}
We present a radiative-convective equilibrium model for young giant exoplanets.  Model predictions are compared with the existing photometric measurements of Planet $\beta$ Pictoris b in the J, H, Ks, L', NB 4.05, M' bands .
 This model will be used to interpret future photometric and spectroscopic observations of exoplanets with SPHERE, mounted at the VLT with a first light expected mid-2014. 
\end{abstract}

\firstsection 

\section{Model}
We developed a radiative-convective equilibrium model for young giant exoplanets. Input parameters are the planet's surface gravity (g), effective temperature (Teff) and elemental composition. Under the additional assumption of thermochemical equilibrium, the model predicts the equilibrium temperature profile and mixing ratio profiles of the most important gases. Opacity sources include the H$_2$-He collision-induced absorption and molecular lines from H$_2$O, CO, CH$_4$, NH$_3$, VO, TiO, Na and K. Line opacity is modeled using k-correlated coefficients pre-calculated over a fixed pressure-temperature grid. Cloud absorption can be added above the expected condensation level (e.g. iron or silicates) with given scale height and optical depth at some reference wavelength. Scattering is not included at the present stage.\\
We compared the model predictions to measurements of planet $\beta$ pictoris b from  \cite[Bonnefoy \etal\ (2013)]{Bonnefoyetal_2013}. 
\section{Results}
We built a grid of models with: a Teff range between 1000 and 2500 K a log$_{10}$(g[cgs])  between 3 and 5, solar system abundances of the elements (\cite[Lodders 2010]{Lodders_2010}) and no cloud opacity.\\
For each model we selected the radius that minimizes the $ \chi^2$ between the observed and calculated apparent magnitudes.
We only kept models with a radius between 0.6 and 2 Jupiter radius (a realistic range derived from evolution models of \cite[Mordasini \etal\ 2012]{Mordasinietal_2012}). \\
The best one (cloud-free model) is a planet with an effective temperature of 1700 K, a log$_{10}$(g[cgs]) of 3 and a radius of $1.53$ $R_{Jup}$ ($ \chi^2= 21$). This model clearly yields too much flux in the  J and H bands (Fig.~1).
\begin{figure}[ht]
\center
\includegraphics[scale=0.35]{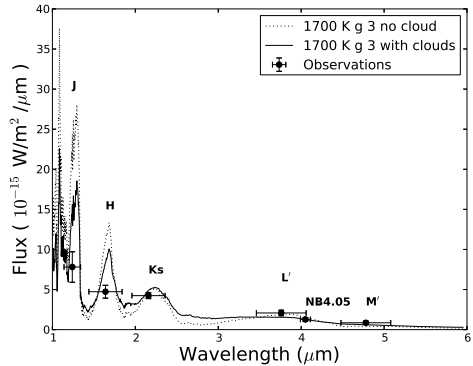} 
\caption{\label{spectre}Apparent flux of our model without (doted line) and with
cloud opacity (solid line), compared with measured apparent
fluxes of $\beta$ pictoris b (circles)
}
\end{figure}\\
In a second step, we added absorption from iron and silicates clouds. We used a particle radius of 30 $\mu$m and assumed the same particle column density for both clouds. For each cloud the opacity is distributed between the condensation level and the 0.1-mbar level with a particle scale height equal to the gas scale height.\\
For Teff = 1700 K and log$_{10}$(g) = 3, using a particulate optical depth ($\tau_{clouds}$) of 0.25 at 1.2 $\mu$m allows us to obtain a $ \chi^2$ of 7.5 for a radius of 1.43 $R_{Jup}$. Compared with the cloud free case, the flux in the J and H bands is lower and that in the Ks, L', NB4.05 and M' bands is higher.
Adding cloud opacity in the model is required to reproduce the data within uncertainties.
\section{Conclusions and perspectives}
In agreement with other models (see references in \cite[Bonnefoy \etal\ 2013]{Bonnefoyetal_2013})  we found that cloud opacity is needed to reproduce the observations of $\beta$ pictoris b.
A model with Teff = 1700 K, log$_{10}$(g[cgs]) = 3, and some cloud opacity agrees with observations within uncertainties, but other combinations of these parameters are probably possible.\\
We plan to explore the parameter space (Teff, g,  $\tau_{clouds}$) for cloudy models of $\beta$ pictoris b, to apply our model to planetary system HR8799, to update methane opacity with the Exomol data base (http://www.exomol.com/), to add NH$_3$ opacity.\\
This model will be used to analyze data from SPHERE  after commissioning on the VLT in 2014.

\vspace{4mm}

\small JLB is founded by the Labex ESEP (http://www.esep.pro/)


\end{document}